\documentclass[12pt]{article}

\usepackage{graphicx}

\def\square{\kern1pt\vbox{\hrule height 1.2pt
\hbox{\vrule width 1.2pt\hskip 3pt
\vbox{\vskip 6pt}\hskip 3pt\vrule width 0.6pt}
\hrule height 0.6pt}\kern1pt}
\def\ltwid{\mathrel{\raise.3ex\hbox{$<$\kern-.75em\lower1ex\hbox{$\sim$}}}}
\def\comp{{\rm C}\llap{\vrule height7.1pt width1pt depth-.4pt\phantom t}}

\def\Fint{\rlap{$\Biggl\rfloor$}\Biggl\lceil}

\begin{document}

\begin{titlepage}
\begin{flushright}
UFIFT-QG-09-04
\end{flushright}

\vspace{0.5cm}

\begin{center}
\bf{A Completely Regular Quantum Stress Tensor with $w < -1$}
\end{center}

\vspace{0.3cm}

\begin{center}
E. O. Kahya$^{\dagger}$ and V. K. Onemli$^{\ddagger}$
\end{center}
\begin{center}
\it{Department of Physics, Ko{\c c} University \\
34450 Sar{\i}yer \.{I}stanbul, TURKEY}
\end{center}

\vspace{0.2cm}

\begin{center}
R. P. Woodard$^{\ast}$
\end{center}
\begin{center}
\it{Department of Physics, University of Florida \\
Gainesville, FL 32611, UNITED STATES}
\end{center}

\vspace{0.3cm}

\begin{center}
ABSTRACT
\end{center}
\hspace{0.3cm} For many quantum field theory computations in cosmology
it is not possible to use the flat space trick of obtaining full,
interacting states by evolving free states over infinite times. State
wave functionals must be specified at finite times and, although the
free states suffice to obtain the lowest order effects, higher order
corrections necessarily involve changes of the initial state. Failing
to correctly change the initial state can result in effective field
equations which diverge on the initial value surface, or which contain
tedious sums of terms that redshift like inverse powers of the scale
factor. In this paper we verify a conjecture from 2004 that the lowest
order initial state correction can indeed absorb the initial value
divergences and all the redshifting terms of the two loop expectation
value of the stress tensor of a massless, minimally coupled scalar
with a quartic self interaction on nondynamical de Sitter background.

\vspace{0.3cm}

\begin{flushleft}
\end{flushleft}

\vspace{0.1cm}

\begin{flushleft}
$^{\dagger}$ e-mail: eokahya@gmail.com \\
$^{\ddagger}$ e-mail: vonemli@ku.edu.tr \\
$^{\ast}$ e-mail: woodard@phys.ufl.edu
\end{flushleft}

\end{titlepage}

\section{Introduction}

Suppose $\varphi(t,\vec{x})$ is a real scalar field operator whose
Lagrangian (by which we mean the spatial integral of the Lagrangian
density) at time $t$ is $L[\varphi(t)]$. Then the relation between the
in-out functional integral formalism and canonical matrix elements is,
\begin{equation}
\Bigl\langle \Phi \Bigl\vert T^*\Bigl(\mathcal{O}[\varphi]\Bigr) \Bigr\vert
\Psi \Bigr\rangle = \Fint [d\phi] \,
e^{i\! \int_{t_1}^{t_2} \! dt \, L[\phi(t)]} \, \Phi^*[\phi(t_2)] \,
\mathcal{O}[\phi] \, \Psi[\phi(t_1)] \; . \label{inout}
\end{equation}
In this formula $\mathcal{O}[\varphi]$ is some functional of the field
for times between $t_1$ and $t_2$, and the $T^*$ symbol means that the
operator upon which it acts is time-ordered, but with any time derivatives
taken {\it outside} the time-ordering. The Heisenberg states $\vert \Psi
\rangle$ and $\vert \Phi \rangle$ have $\comp$-number wave functionals
$\Psi[\phi(t_1)]$ and $\Phi[\phi(t_2)]$ in terms of the eigenkets of
$\varphi(t,\vec{x})$ at times $t_1$ and $t_2$, respectively.

In flat space physics we typically seek to compute matrix elements
between states which are true vacuum in the infinite past and future.
This might seem problematic because no one has ever exhibited a
normalizable energy eigenstate for an interacting, $D=4$ dimensional
quantum field theory. Of course it would be possible to build up
perturbative corrections --- which is all that is needed for finite
order computations --- the same as in quantum mechanics. However, for
theories with a mass gap we can avoid this tedious and noncovariant
exercise by taking $\vert \Psi \rangle$ and $\vert \Phi \rangle$ to be
free vacuum, and then considering the limit in which $t_1$ goes to
$-\infty$ and $t_2$ goes to $+\infty$. Up to a normalization factor,
this limit projects out true vacuum in the weak operator sense \cite{IZ}.
Of course the most interesting theories have massless particles, which
violate the assumption about a mass gap, but it is believed the
procedure still gives correct inclusive rates and cross sections
\cite{Weinberg}.

In cosmology we typically imagine that the universe began with an
initial singularity, and it is often our ignorance about what happens
in the far future that is the chief reason for interest in the
computation. The canonical operator formalism is of course the same,
but its more useful functional integral representation is given by
the Schwinger-Keldysh formalism \cite{JS,M,K}. The relation analogous
to (\ref{inout}) is \cite{FW},
\begin{eqnarray}
\lefteqn{\Bigl\langle \Psi \Bigl\vert \overline{T}^*\Bigl(B[\varphi]\Bigr)
T^*\Bigl(A[\varphi]\Bigr) \Bigr\vert \Psi \Bigr\rangle
= \Fint [d\phi_{\scriptscriptstyle +}] [d\phi_{\scriptscriptstyle -}] \,
\delta\Bigl[\phi_{\scriptscriptstyle -}\!(t_2) \!-\! \phi_{\scriptscriptstyle
+}\!(t_2)\Bigr] } \nonumber \\
& & \hspace{2.3cm} \times e^{i \! \int_{t_1}^{t_2} \! dt \, \Bigl\{
L[\phi_{\scriptscriptstyle +}\!(t)] \!-\! L[\phi_{\scriptscriptstyle -}\!(t)]
\Bigr\}} \, \Psi^*[\phi_{\scriptscriptstyle -}\!(t_1)]
B[\phi_{\scriptscriptstyle -}] A[\phi_{\scriptscriptstyle +}]
\Psi[\phi_{\scriptscriptstyle +}\!(t_1)] \; . \qquad \label{fund}
\end{eqnarray}
Here $\vert \Psi\rangle$ is the Heisenberg state whose $\comp$-number wave
functional in terms of the time $t_1$ eigenkets is $\Psi[\phi(t_1)]$. Like
$\mathcal{O}[\varphi]$ in (\ref{inout}), the operators $A[\varphi]$ and
$B[\varphi]$ are functionals of the operators $\varphi(t,\vec{x})$ for
$t_1 < t < t_2$. As in (\ref{inout}), the $T^*$ symbol stands for
time-ordering, with any time derivatives taken outside; the $\overline{T}^*$
symbol stands for anti-time-ordering, again with time derivatives taken
outside. The reason for the two $\comp$-number integration variables $\phi_{
\scriptscriptstyle \pm}\!(t,\vec{x})$ in (\ref{fund}) is that the functional
integration over $\phi_{\scriptscriptstyle +}$ evolves the system forward
to time $t_2$, whereas the functional integration over $\phi_{
\scriptscriptstyle -}$ carries it back to the initial time $t_1$.

Expression (\ref{fund}) is well adapted to cosmological problems in which
the universe is released in a prepared state $\vert \Psi\rangle$ at some
finite time $t_1$ and its subsequent evolution is studied through
correlators. Unfortunately, we can no longer use infinite time evolution
to transform the known free states into fully interacting ones. There
have been attempts to achieve the same thing by including an additional
evolution in Euclidean time \cite{Hartle,Maldacena,Polyakov}. However, the
absence of a unique vacuum means that it is not clear what the fully
interacting state should be \cite{BD}. Of special significance to this
work is the fact that this is even true on de Sitter background for the
massless, minimally coupled scalar \cite{AF}.

Of course we are doing perturbation theory so the lowest order results
can be obtained using the free vacuum. Certain higher order corrections
show secular growth from the coherent superposition of interactions
throughout the past light-cone, which is not affected by corrections to
the initial state \cite{TW1,OW,SQED1,Yukawa,QGDirac,SQED2,KO,TW2}. However,
in many cases state corrections on the initial value surface are as
important as 4-volume effects \cite{KW1,DW,KW2}. And even when a higher
order correction is dominated by secular growth from a 4-volume effect,
failure to include state corrections leads to a number of problems
including:
\begin{itemize}
\item{Divergences when operators touch the initial value surface
\cite{OW,SQED1,Yukawa,QGDirac,KO,KW1,DW,KW2,BOW};}
\item{Nonvanishing surface terms from partial integrations
\cite{SQED2}; and} \item{Complicated collections of terms which
redshift like inverse powers of the scale factor \cite{OW}.}
\end{itemize}

This paper concerns an example of the first and last problems above.
Consider a massless, minimally coupled scalar with a $\lambda \varphi^4$
self interaction on nondynamical de Sitter background whose scale factor
$a = e^{Ht}$ is normalized to be one on the initial value surface. The
expectation value of the stress tensor has been computed at one and two loop
orders in the presence of free Bunch-Davies vacuum \cite{OW}. With a slight
change in the original renormalization scheme, the energy density and pressure
are \cite{RGF},
\begin{eqnarray}
\lefteqn{\rho = \frac{3 H^2}{8 \pi G} + \frac{\lambda H^4}{(4\pi)^4} \Biggl\{
+2 \ln^2(a) + \frac{13}6 \ln(a) -\frac{43}{18} + \frac{\pi^2}3 } \nonumber \\
& & \hspace{6.3cm} + \frac{8}{9 a^3} - 2 \sum_{n=2}^{\infty}
\frac{(n \!+\! 1)}{n^2 a^n} \Biggr\} + O(\lambda^2) \; , \qquad \label{rho} \\
\lefteqn{p = -\frac{3 H^2}{8 \pi G} + \frac{\lambda H^4}{(4\pi)^4} \Biggl\{-2
\ln^2(a) -\frac72 \ln(a) + \frac53 -\frac{\pi^2}3 } \nonumber \\
& & \hspace{6.3cm} -\frac23 \sum_{n=2}^{\infty} \frac{(n\!-\!3)
(n\!+\!1)}{n^2 a^n} \Biggr\} + O(\lambda^2) \; . \label{pres}
\end{eqnarray}
The model was particularly curious to us since it leads to
$p/\rho\equiv w<-1$ which has been a main area of interest in
recent years \cite{refs}. The secular growth in
(\ref{rho}-\ref{pres}) derives from inflationary particle
production driving the scalar field strength up its $\lambda
\varphi^4$ potential, which of course increases the vacuum
energy.\footnote{ Because $\lambda$ is a constant, whereas $\ln(a)
= H t$ grows with time, these secular corrections eventually
become nonperturbatively strong. Starobinsky has developed a
stochastic formalism for summing the series of leading logarithms
\cite{AAS,SY,TW3}.} This part of the result will persist for any
initial state which is finitely excited from Bunch-Davies vacuum.
That is not true of the exponentially falling terms,
\begin{eqnarray}
\rho_{\rm falling} & = & \frac{\lambda H^4}{(4\pi)^4} \Biggl\{-\frac3{2 a^2}
-2 \sum_{n=4}^{\infty} \frac{(n\!+\!1)}{n^2 a^n} \Biggr\} \; , \label{rfall} \\
p_{\rm falling} & = & \frac{\lambda H^4}{(4\pi)^4} \Biggl\{+\frac1{2 a^2}
-\frac23 \sum_{n=4}^{\infty} \frac{(n\!-\!3)(n\!+\!1)}{n^2 a^n} \Biggr\}
\; . \qquad \label{pfall}
\end{eqnarray}
Because they are separately conserved, diverge on the initial value surface,
and fall off rapidly as one evolves to late times, it was conjectured that
$\rho_{\rm falling}$ and $p_{\rm falling}$ could be absorbed into corrections
to the initial state wave functional \cite{OW}. In this paper we will prove the
conjecture by constructing the $\lambda \phi^2$ correction which completely
absorbs (\ref{rfall}-\ref{pfall}). We will even explain the curious fact that
they contain no $1/a^3$ term.

This paper consists of five sections of which the first is ending. In section
2 we specify the background geometry and the entire apparatus of perturbation
theory, even though our own work does not require regularization,
renormalization or even the quartic self-interaction. In section 3 we compute
the effect on the expectation value of the stress tensor of a general
$\lambda \phi^2$ correction to the initial state wave functionals. The
specific correction which absorbs (\ref{rfall}-\ref{pfall}) is worked out
in section 4. Our conclusions are given in section 5.

\section{$\lambda \varphi^4$ Theory on de Sitter}

We work on the open conformal coordinate patch of de Sitter space, the
invariant element for which is,
\begin{equation}
ds^2 \equiv g_{\mu\nu} dx^{\mu} dx^{\nu} = a^2 \Bigl[-d\eta^2 + d\vec{x} \cdot
d\vec{x}\Bigr] \qquad {\rm with} \qquad a \equiv -\frac1{H \eta} = e^{Ht} \; .
\end{equation}
The Hubble constant is $H$ and the conformal time $\eta$ runs from
$-\infty$ to $0$. To facilitate dimensional regularization (when
necessary) we work in $D$ spacetime dimensions, with the indices
taking values $\mu,\nu = 0,1,2, \dots,(D\!-\!1)$. As the name of
the coordinate patch suggests, the metric is conformal to the flat
space metric $\eta_{\mu\nu}$: $g_{\mu\nu} = a^2 \eta_{\mu\nu}$. It
is sometimes useful to distinguish the purely spatial parts of
tensors with an overline, for example,
\begin{equation}
\langle \Omega \vert T_{\mu\nu} \vert \Omega \rangle \equiv a^2 \delta^0_{\mu}
\delta^0_{\nu} \times \rho + a^2 \overline{\eta}_{\mu\nu} \times p \; .
\end{equation}

The Lagrangian density is,
\begin{equation}
\mathcal{L} = -\frac12 \partial_{\mu} \varphi_0 \partial_{\nu}
\varphi_0 g^{\mu\nu} \sqrt{-g} -\frac{\xi_0}2 \varphi_0^2 R \sqrt{-g}
-\frac{\lambda_0}{4!} \varphi_0^4 \sqrt{-g} -
\frac{(D\!-\!2) \Lambda_0}{16 \pi G} \, \sqrt{-g} \; .
\end{equation}
Here $\varphi_0$ is the bare field, $\xi_0$ is the bare conformal coupling
constant, $\lambda_0$ is the bare quartic coupling constant, and $\Lambda_0$
is the bare cosmological constant. The renormalized field $\varphi$ is
defined by field strength renormalization of the bare one as usual,
\begin{equation}
\varphi(x) \equiv \frac1{\sqrt{Z}} \, \varphi_0(x) \; .
\end{equation}
That brings the Lagrangian density to the form,
\begin{equation}
\mathcal{L} = -\frac{Z}2 \partial_{\mu} \varphi \partial_{\nu}
\varphi g^{\mu\nu} \sqrt{-g} -\frac{Z \xi_0}2 \varphi^2 R \sqrt{-g}
-\frac{Z^2 \lambda_0}{4!} \varphi^4 \sqrt{-g} -
\frac{(D\!-\!2) \Lambda_0}{16 \pi G} \, \sqrt{-g} \; .
\end{equation}
The associated stress tensor is,
\begin{eqnarray}
\lefteqn{T_{\mu\nu} = Z\Bigl[\delta^{\rho}_{\mu} \delta^{\sigma}_{\nu}
-\frac12 g_{\mu\nu} g^{\rho\sigma}\Bigr] \partial_{\rho} \varphi
\partial_{\sigma} \varphi -\frac{Z^2 \lambda_0}{4!} \varphi^4 g_{\mu\nu} }
\nonumber \\
& & \hspace{2.3cm} + Z\xi_0 \Bigl[ R_{\mu\nu} \!-\! \frac12
g_{\mu\nu} R \!-\! D_{\mu} D_{\nu} \!+\! g_{\mu\nu} \square\Bigr]
\varphi^2 - \frac{(D\!-\!2) \Lambda_0}{16 \pi G} \, g_{\mu\nu} \;
. \qquad
\end{eqnarray}
Conservation is straightforward to verify, as a strong operator equation,
using the regulated scalar field equation,
\begin{equation}
T_{\mu\nu ;}^{~~~ \nu} = \Biggl[ Z \square \varphi - Z \xi_0 R \varphi
- \frac{Z^2 \lambda_0}{6} \, \varphi^3\Biggr] \partial_{\mu} \varphi = 0
\; . \label{conser}
\end{equation}

Renormalization is accomplished by expressing the bare parameters in terms
of the renormalized parameters and counter parameters,
\begin{equation}
Z \equiv 1 + \delta Z \quad , \quad
Z^2 \lambda_0 \equiv \lambda + \delta \lambda \quad , \quad
Z \xi_0 \equiv 0 + \delta \xi \quad , \quad
\Lambda_0 \equiv \frac{6 H^2}{D\!-\!2} + \delta \Lambda \; .
\end{equation}
Note that no mass counterterm is necessary because mass is multiplicatively
renormalized in dimensional regularization. However, a conformal counterterm
is necessary even if the renormalized conformal coupling is zero. The one and
two loop counterterms were chosen as the following functions of $\epsilon
\equiv 4 - D$,
\begin{eqnarray}
\delta Z & = & -\frac{\lambda^2}{12 (4\pi)^4} \Bigl(\frac{4\pi}{H^2}\Bigr)^{\!
\epsilon} \frac{\Gamma^2(1 \!-\! \frac12 \epsilon)}{(1 \!-\! \frac32 \epsilon)
(1 \!-\! \epsilon) (1 \!-\! \frac34 \epsilon) \epsilon} + O(\lambda^3) \; , \\
\delta \lambda & = & \frac{3 \lambda^2}{16 \pi^2} \Bigl(\frac{4\pi}{H^2}
\Bigr)^{\!\frac12 \epsilon} \frac{\Gamma(1 \!-\! \frac12 \epsilon)}{(1 \!-\!
\epsilon) \epsilon} + O(\lambda^2) \; , \\
\delta \xi & = & -\frac{\lambda}{192 \pi^2} \Bigl(\frac{4\pi}{H^2}\Bigr)^{\!
\frac12 \epsilon} \frac{\pi \cot(\frac12 \pi \epsilon) (1 \!-\! \epsilon)
\Gamma(1 \!-\! \epsilon)}{(1 \!-\! \frac13 \epsilon) (1\!-\! \frac14 \epsilon)
\Gamma(1 \!-\! \frac12 \epsilon)} + O(\lambda^2) \; , \label{dxi} \\
\delta \Lambda & = & \frac{8 \pi G H^4}{D \!-\!2} \Biggl\{ \frac3{16\pi^2}
\Bigl(\frac{4\pi}{H^2}\Bigr)^{\!\frac12 \epsilon} \frac{(1 \!-\! \epsilon)
(1 \!-\! \frac12 \epsilon) (1 \!-\! \frac13 \epsilon) \Gamma(1\!-\!\epsilon)}{
(1 \!-\! \frac14 \epsilon) \Gamma(1 \!-\! \frac12 \epsilon)} \nonumber \\
& & \hspace{2cm} - \frac{\lambda}{(4\pi)^4} \Bigl(\frac{4\pi}{H^2}\Bigr)^{\!
\epsilon} \frac{[\pi \cot(\frac12 \pi \epsilon) \epsilon (1 \!-\! \epsilon)
\Gamma(1 \!-\! \epsilon)]^2}{4 \epsilon (1 \!-\! \frac14 \epsilon)
\Gamma^2(1 \!-\! \frac12 \epsilon)} + O(\lambda^2) \Biggr\} . \qquad
\end{eqnarray}
Note that a more complicated renormalization scheme involving a mass
counterterm was employed in the original computation \cite{OW}.

There are no normalizable de Sitter invariant state for the free massless,
minimally coupled scalar \cite{AF}. We choose to preserve the symmetries
of cosmology --- homogeneity and isotropy --- which is known as the ``E3''
vacuum \cite{BA}. It can be realized in terms of plane wave mode sums by
making the spatial manifold $T^{D-1}$, rather than $R^{D-1}$, with coordinate
radius $H^{-1}$ in each direction, and then using the integral approximation
with the lower limit cut off at $k=H$ \cite{TW4,ITTW,JMPW}. The resulting
free field expansion is,
\begin{equation}
\varphi(\eta,\vec{x}) = \int \!\! \frac{d^{D-1}k}{(2\pi)^{D-1}} \,
\theta(k\!-\!H) \Biggl\{ u(\eta,k) e^{i\vec{k} \cdot \vec{x}} \alpha(\vec{k})
+ u^*(\eta,k) e^{-i\vec{k} \cdot \vec{x}} \alpha^{\dagger}\!(\vec{k})
\Biggr\} . \label{free}
\end{equation}
In this expression the creation and annihilation operators are canonically
normalized,
\begin{equation}
\Bigl[\alpha(\vec{k}) , \alpha^{\dagger}\!(\vec{k}')\Bigr] = (2\pi)^{D-1}
\delta^{D-1}\!(\vec{k} \!-\! \vec{k}') \; ,
\end{equation}
and the mode functions are,
\begin{equation}
u(\eta,k) = \sqrt{\frac{\pi}{4H}} a^{-\frac{D-1}2} H^{(1)}_{\frac{D-1}2}\!
\Bigl(\frac{k}{H a}\Bigr) \; . \label{udef}
\end{equation}
The mode functions take a particularly simple for in $D=4$,
\begin{equation}
u(\eta,k) \Bigl\vert_{D=4} = \frac{H}{\sqrt{2 k^3}} \Bigl[1 - \frac{ik}{H a}
\Bigr] \exp\Bigl[\frac{ik}{H a}\Bigr] \; . \label{u4}
\end{equation}

Because time translation is not an invariance of cosmology there is no
conserved energy, even at the free level. However, it is still the case
that each mode of a free quantum field theory behaves as a harmonic
oscillator, in this case with time dependent mass and frequency. Hence
there will be a minimum energy Heisenberg state at any instant, although
this state will not generally have the minimum energy before or after
that instant. Bunch-Davies vacuum is the state which was minimum energy
in the distant past. It corresponds to the condition,
\begin{equation}
\alpha(\vec{k}) \Bigl\vert \Omega \Bigr\rangle = 0 \qquad \forall \vec{k}
\ni \Vert \vec{k}\Vert > H \; . \label{Bunch}
\end{equation}
It is a straightforward exercise to solve for the state wave
functional using expressions (\ref{free}), (\ref{udef}) and
(\ref{Bunch}),
\begin{equation}
\Omega\Bigl[\phi(\eta_I)\Bigr] = N \exp\Biggl[-\frac12 \! \int \!\!
\frac{d^{D-1}k}{(2\pi)^{D-1}} \, \theta(k \!-\! H) \widetilde{\phi}^*(\eta_I,
\vec{k}) \Bigl[\frac{i u'(\eta_I,k)}{u(\eta_I,k)}\Bigr]^*
\widetilde{\phi}(\eta_I,\vec{k}) \Biggr] \; . \label{freestate}
\end{equation}
Here $\eta_I \equiv -1/H$ is the initial time (corresponding to $t=0$),
$N$ is a functional normalization factor and $\widetilde{\phi}(\eta_I,\vec{k})$
is the spatial Fourier transform of field on the initial value surface,
\begin{equation}
\widetilde{\phi}(\eta_I,\vec{k}) \equiv \int \! d^{D-1}x \, e^{-i\vec{k} \cdot
\vec{x}} \phi(\eta_I,\vec{x}) \; .
\end{equation}

It remains only to give the Schwinger-Keldysh formalism, which can be
read off from the fundamental relation (\ref{fund}). There are some
excellent reviews of this subject \cite{revs} so we shall just summarize
the results:
\begin{itemize}
\item{Because the same field operator $\varphi(\eta,\vec{x})$ is
represented by two different functional integration variables $\phi_{
\scriptscriptstyle \pm}(\eta,\vec{x})$, the endpoints of lines carry a
$\pm$ polarity;}
\item{Interaction vertices are either all $+$ or all $-$;}
\item{Vertices with a $+$ polarity are the same as for the in-out
formalism whereas those with a $-$ polarity are conjugated;}
\item{Corrections to the initial states take the form of vertices on
the initial value surface; and}
\item{Propagators can be $++$, $+-$, $-+$ or $--$.}
\end{itemize}
The mode sums for the various propagators are,
\begin{eqnarray}
i\Delta_{\scriptscriptstyle ++}\!(x;x') & \!\!\!=\!\!\!& \int \!\!
\frac{d^{D-1}k}{(2\pi)^{D-1}} \theta(k \!-\! H) e^{i\vec{k} \cdot (\vec{x}
- \vec{x}')} \nonumber \\
& & \hspace{.5cm} \times \Biggl\{ \theta(\eta \!-\! \eta') u(\eta,k)
u^*(\eta',k) + \theta(\eta' \!-\! \eta) u^*(\eta,k) u(\eta',k) \Biggr\} , \\
i\Delta_{\scriptscriptstyle +-}\!(x;x') & \!\!\!=\!\!\!& \int \!\!
\frac{d^{D-1}k}{(2\pi)^{D-1}} \theta(k \!-\! H) e^{i\vec{k} \cdot (\vec{x}
- \vec{x}')} u^*(\eta,k) u(\eta',k) \; , \qquad \\
i\Delta_{\scriptscriptstyle -+}\!(x;x') & \!\!\!=\!\!\!& \int \!\!
\frac{d^{D-1}k}{(2\pi)^{D-1}} \theta(k \!-\! H) e^{i\vec{k} \cdot (\vec{x}
- \vec{x}')} u(\eta,k) u^*(\eta',k) \; , \qquad \\
i\Delta_{\scriptscriptstyle --}\!(x;x') & \!\!\!=\!\!\! & \int \!\!
\frac{d^{D-1}k}{(2\pi)^{D-1}} \theta(k \!-\! H) e^{i\vec{k} \cdot (\vec{x}
- \vec{x}')} \nonumber \\
& & \hspace{.5cm} \times \Biggl\{ \theta(\eta \!-\! \eta') u^*(\eta,k)
u(\eta',k) + \theta(\eta' \!-\! \eta) u(\eta,k) u^*(\eta',k) \Biggr\} . \qquad
\end{eqnarray}

\section{Order $\lambda \varphi^2$ State Correction}

Consider a change in the initial state,
\begin{equation}
\Bigl\vert \Omega \Bigl\rangle \longrightarrow \Bigl\vert \Psi \Bigl\rangle
\equiv \Bigl\vert \Omega \Bigl\rangle + \Bigl\vert \Delta \Omega \Bigl\rangle
\; .
\end{equation}
where $\vert \Omega \rangle$ is free, Bunch-Davies vacuum
(\ref{freestate}). Because the stress tensor is conserved
(\ref{conser}) as a strong operator equation, its expectation
value must be conserved in any state. Hence we have,
\begin{equation}
D^{\nu} \Bigl\langle \Omega \Bigl\vert T_{\mu\nu} \Bigr\vert \Omega
\Bigr\rangle = 0 \; ,
\end{equation}
and also,
\begin{equation}
D^{\nu} \Biggl\{ \Bigl\langle \Delta \Omega \Bigl\vert T_{\mu\nu} \Bigr\vert
\Omega \Bigr\rangle + \Bigl\langle \Omega \Bigl\vert T_{\mu\nu} \Bigr\vert
\Delta \Omega \Bigr\rangle + \Bigl\langle \Delta \Omega \Bigl\vert T_{\mu\nu}
\Bigr\vert \Delta \Omega \Bigr\rangle \Biggr\} = 0 \; .
\end{equation}
Of course this was one reason for suspecting that the separately conserved
parts (\ref{rfall}-\ref{pfall}) of the original result (\ref{rho}-\ref{pres})
could be absorbed into a change of the initial state.

\begin{figure}
\begin{center}
\includegraphics[width=5in]{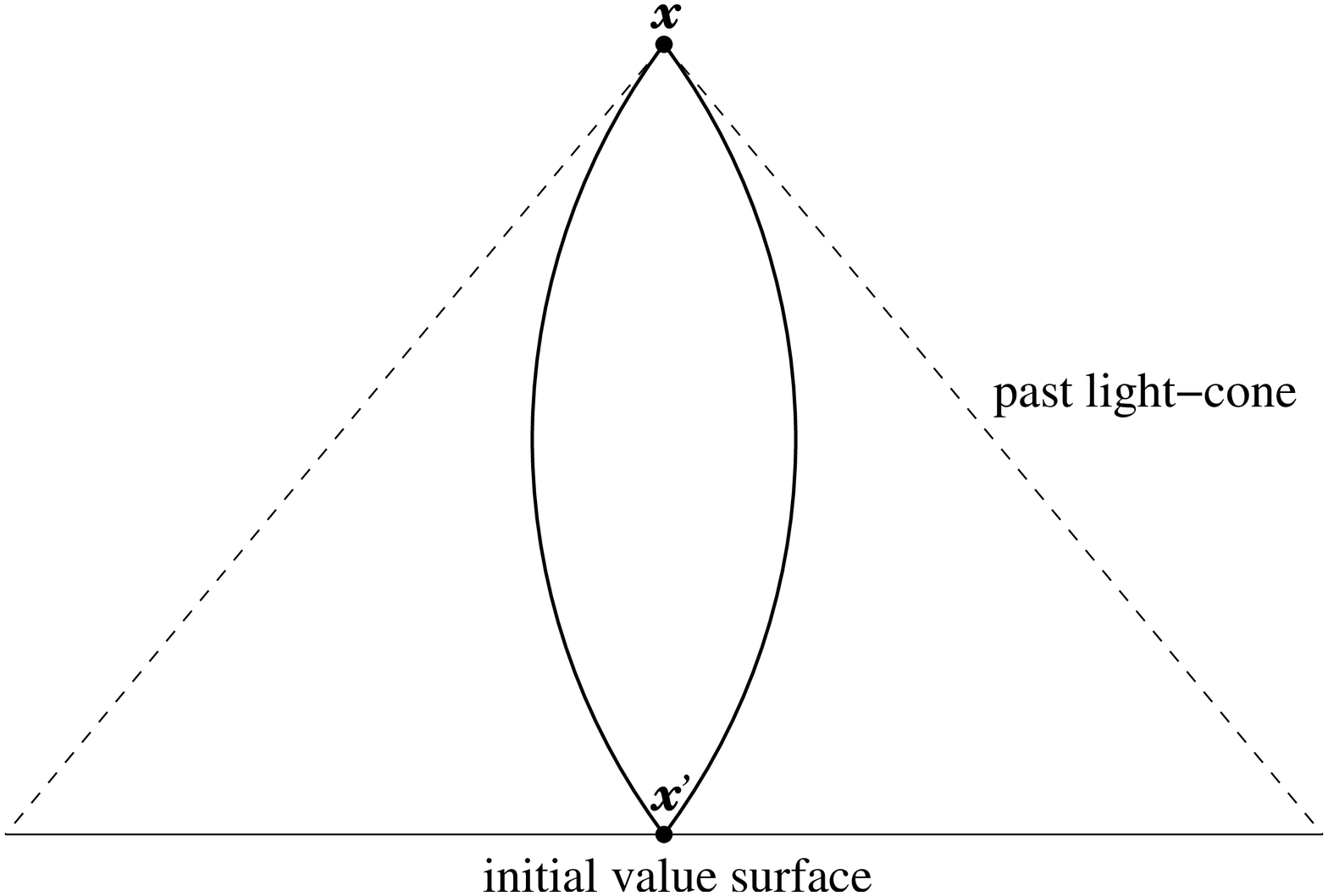}
\end{center}
\caption{Feynman diagram for the order $\lambda$ intial state correction to
the expectation value of the stress tensor at $x^{\mu}$.}
\label{graph}
\end{figure}

The expectation value of the stress tensor must also be conserved
order-by-order in perturbations theory. Of course we can expand
the initial state correction in powers of $\lambda$,
\begin{equation}
\Bigl\vert \Delta \Omega \Bigr\rangle \equiv \sum_{n=1}^{\infty} \lambda^n
\Bigl\vert \Omega_n\Bigr\rangle \; .
\end{equation}
The purpose of this paper is to find the first order correction $\lambda
\vert \Omega_1 \rangle$ which absorbs the exponentially redshifting terms
(\ref{rfall}-\ref{pfall}),
\begin{eqnarray}
\lefteqn{\lambda \Bigl[\delta^{\rho}_{\mu} \delta^{\sigma}_{\nu} - \frac12
g_{\mu\nu} g^{\rho\sigma}\Bigr] \Biggl\{ \Bigl\langle \Omega_1 \Bigl\vert
\partial_{\rho} \varphi \partial_{\sigma} \varphi \Bigr\vert \Omega
\Bigr\rangle + \Bigl\langle \Omega \Bigl\vert \partial_{\rho} \varphi
\partial_{\sigma} \varphi \Bigr\vert \Omega_1 \Bigr\rangle \Biggr\} }
\nonumber \\
& & \hspace{5.4cm} = - a^2 \delta^0_{\mu} \delta^0_{\nu} \times
\rho_{\rm falling} - a^2 \overline{\eta}_{\mu\nu} \times p_{\rm falling}
\; . \qquad
\end{eqnarray}

In order for perturbation theory to make sense, all corrections to
the initial state must take the form of the free vacuum times
powers of the fields. Because the first order correction of
interest to us must link up with the $\partial_{\rho} \varphi
\partial_{\sigma} \varphi$ part of the stress tensor, we are
obviously looking for a correction of the form $\lambda \phi^2$.
The two fields in the state correction will each connect with
fields in the stress tensor as in Fig.~\ref{graph}, so there will
be no ultraviolet divergences and we can simplify the discussion
by taking $D=4$. The most general state correction with these
properties, which also has the right dimensions and is consistent
with homogeneity and isotropy, can be written as,
\begin{eqnarray}
\lambda \Omega_1\Bigl[\phi_{\scriptscriptstyle +}(\eta_I)\Bigr] &
= & \Omega\Bigl[\phi_{\scriptscriptstyle +}(\eta_I)\Bigr] \times
\frac{\lambda H}{2} \int \!\! \frac{d^3k}{(2\pi)^3} \,
F\Bigl(\frac{k}{H}\Bigr) \widetilde{\phi}_{\scriptscriptstyle
+}^*(\eta_I,\vec{k}) \widetilde{\phi}_{\scriptscriptstyle
+}(\eta_I,\vec{k}) \; ,
\qquad \label{+cor} \\
\lambda \Omega_1^*\Bigl[\phi_{\scriptscriptstyle -}(\eta_I)\Bigr]
& = & \Omega\Bigl[\phi_{\scriptscriptstyle -}(\eta_I)\Bigr] \times
\frac{\lambda H}{2} \int \!\! \frac{d^3k}{(2\pi)^3} \,
F^*\Bigl(\frac{k}{H}\Bigr) \widetilde{\phi}_{\scriptscriptstyle
-}^*(\eta_I,\vec{k}) \widetilde{\phi}_{\scriptscriptstyle
-}(\eta_I,\vec{k}) \; . \qquad \label{-cor}
\end{eqnarray}
The function $F(k/H)$ characterizes the state, and is at this stage
arbitrary. We will determine it in the next section.

State corrections of the form (\ref{+cor}-\ref{-cor}) are treated
just as interaction vertices in the Schwinger-Keldysh formalism,
the only differences with the volume terms being that there is no
factor of $\pm i$, that the ``interactions'' are restricted to the
initial value surface, and that they are generally not local in
position space. We obviously get distinct contributions from the
$\phi_{\scriptscriptstyle +}$ correction (\ref{+cor}) and from the
$\phi_{\scriptscriptstyle -}$ correction (\ref{-cor}). The
contribution from (\ref{+cor}) involves two $++$ propagators
between the observation point $(\eta,\vec{x})$ and the initial
value surface. Because the observation comes after the initial
time this contribution is,
\begin{eqnarray}
\lefteqn{\Delta T_{\mu\nu}^+ = \lambda H \Bigl[\delta^{\rho}_{\mu}
\delta^{\sigma}_{\nu} - \frac12 g_{\mu\nu} g^{\rho\sigma}\Bigr]
\int \!\! \frac{d^3k}{(2\pi)^3} \, F\Bigl(\frac{k}{H}\Bigr) \Bigl[
u^*(\eta_I,k)\Bigr]^2 }
\nonumber \\
& & \hspace{6cm} \times \partial_{\rho} \Bigl[ e^{i \vec{k} \cdot \vec{x}}
u(\eta,k)\Bigr] \partial_{\sigma} \Bigl[ e^{-i \vec{k} \cdot \vec{x}}
u(\eta,k)\Bigr] \; . \qquad \label{DT+}
\end{eqnarray}
The contribution from (\ref{-cor}) involves two $+-$ propagators and is,
\begin{eqnarray}
\lefteqn{\Delta T_{\mu\nu}^- = \lambda H \Bigl[\delta^{\rho}_{\mu}
\delta^{\sigma}_{\nu} - \frac12 g_{\mu\nu} g^{\rho\sigma}\Bigr]
\int \!\! \frac{d^3k}{(2\pi)^3} \, F^*\Bigl(\frac{k}{H}\Bigr)
\Bigl[ u(\eta_I,k)\Bigr]^2 }
\nonumber \\
& & \hspace{5.5cm} \times \partial_{\rho} \Bigl[ e^{i \vec{k} \cdot \vec{x}}
u^*(\eta,k)\Bigr] \partial_{\sigma} \Bigl[ e^{-i \vec{k} \cdot \vec{x}}
u^*(\eta,k)\Bigr] \; . \qquad \label{DT-}
\end{eqnarray}
We obviously wish to solve for $F(k/H)$ to enforce the condition,
\begin{equation}
\Delta T_{\mu\nu}^+ + \Delta T_{\mu\nu}^- = - a^2 \delta^0_{\mu}
\delta^0_{\nu} \times \rho_{\rm falling} - a^2 \overline{\eta}_{\mu\nu}
\times p_{\rm falling} \; . \label{condition}
\end{equation}

We can eliminate the tensor algebra by distinguishing the temporal and
spatial derivatives,
\begin{eqnarray}
A & \equiv & \lambda H \int \!\! \frac{d^3k}{(2\pi)^3} \, F\Bigl(\frac{k}{H}
\Bigr) \Bigl[ u^*(\eta_I,k)\Bigr]^2 \Bigl[ \partial_0 u(\eta,k)\Bigr]^2 \; , \\
& = & \frac{\lambda H a^2}{2\pi^2} \int_0^{\infty} \!\! dk \, k^2 F\Bigl(
\frac{k}{H}\Bigr) \Bigl[ u^*(\eta_I,k)\Bigr]^2 \Bigl[\frac1{a} \partial_0
u(\eta,k)\Bigr]^2 \; , \\
B & \equiv & \lambda H \int \!\! \frac{d^3k}{(2\pi)^3} \, F\Bigl(\frac{k}{H}
\Bigr) \Bigl[ u^*(\eta_I,k)\Bigr]^2 \Bigl[ k u(\eta,k)\Bigr]^2 \; , \\
& = & \frac{\lambda H a^2}{2\pi^2} \int_0^{\infty} \!\! dk \, k^2 F\Bigl(
\frac{k}{H}\Bigr) \Bigl[ u^*(\eta_I,k)\Bigr]^2 \Bigl[\frac{k}{a} u(\eta,k)
\Bigr]^2 \; .
\end{eqnarray}
Decomposing $\Delta T^{+}_{\mu\nu}$ into its induced energy density
and pressure gives,
\begin{eqnarray}
\Delta \rho^+ &\!\!\! = \!\!\! & \frac1{a^2} \Bigl[A + B\Bigr] \; , \\
&\!\!\! = \!\!\!& \frac{\lambda H}{4\pi^2} \! \int_0^{\infty} \!\!\!\! dk \,k^2
\, F\Bigl(\frac{k}{H}\Bigr) \Bigl[u^*(\eta_I,k)\Bigr]^2 \Biggl\{ \Bigl[
\frac1{a} \partial_0 u(\eta,k)\Bigr]^2 + \Bigl[ \frac{k}{a} u(\eta,k)\Bigr]^2
\Biggr\} \; , \label{rho+} \qquad \\
\Delta p^+ & \!\!\!=\!\!\! & \frac1{2 a^2} \Bigl[A - \frac13 B\Bigr] \; , \\
&\!\!\! =\!\!\! & \frac{\lambda H}{4\pi^2} \! \int_0^{\infty} \!\!\!\! dk \,k^2
\, F\Bigl(\frac{k}{H}\Bigr) \Bigl[u^*(\eta_I,k)\Bigr]^2 \Biggl\{ \Bigl[
\frac1{a} \partial_0 u(\eta,k)\Bigr]^2 - \frac13 \Bigl[ \frac{k}{a} u(\eta,k)
\Bigr]^2 \Biggr\} \; . \qquad
\end{eqnarray}
The $-$ contributions follow from complex conjugation, and we will be able
to completely absorb the exponentially falling terms (\ref{rfall}-\ref{pfall})
if the function $F(k/H)$ can be chosen such that,
\begin{equation}
\Delta \rho^+ + (\Delta \rho^+)^* = -\rho_{\rm falling} \qquad {\rm and}
\qquad \Delta p^+ + (\Delta p^+)^* = -p_{\rm falling} \; . \label{3fin}
\end{equation}

\section{Reconstructing $F(k/H)$}

Because the stress tensor is conserved it suffices to enforce just the first
condition of (\ref{3fin}). The key to doing this is expanding out the
exponentially falling terms in the curly brackets of expression (\ref{rho+}).
As we saw in expression (\ref{u4}) the mode function and its time derivative
are simple in $D=4$ spacetime dimensions,
\begin{equation}
u(\eta,k) = \frac{H}{\sqrt{2 k^3}} \Bigl[1 \!-\!
\frac{ik}{Ha}\Bigr] \exp\Bigl[\frac{ik}{Ha}\Bigr] \;
\Longrightarrow \; \partial_0 u(\eta,k) = \frac{H}{\sqrt{2 k^3}}
\Bigl[ -\frac{k^2}{H a}\Bigr] \exp\Bigl[\frac{ik}{Ha}\Bigr] \; .
\end{equation}
It follows that the curly bracketed term of (\ref{rho+}) is,
\begin{eqnarray}
\Bigl[\frac1{a} \partial_0 u(\eta,k)\Bigr]^2 + \Bigl[ \frac{k}{a}
u(\eta,k)\Bigr]^2 & = & \frac{H^4}{2 k^3} \Bigl(\frac{k}{Ha}\Bigr)^2
\Bigl[1 - \frac{2i k}{H a}\Bigr] \exp\Bigl[\frac{2ik}{Ha}\Bigr] \; , \qquad \\
& = & \frac{H^4}{2 k^3} \Bigl(\frac{k}{Ha}\Bigr)^2 \Biggl\{1 -
\sum_{n=2}^{\infty} \frac{(n \!-\!1)}{n!} \Bigl(\frac{2i k}{H a}\Bigr)^n
\Biggr\} \; . \qquad \label{series}
\end{eqnarray}
It is immediately apparent why there are no $1/a^3$ terms in $\rho_{\rm
falling}$!

Substituting (\ref{series}) into expression (\ref{rho+}) and making the
change of variable $k = H x$ gives,
\begin{equation}
\Delta \rho^+ = \frac{\lambda H^4}{16 \pi^2} \int_0^{\infty} \!\!\! dx \,
\frac{F(x)}{x^2} \, (1 \!+\! ix)^2 e^{-2i x} \Biggl\{ \frac1{a^2} -
\sum_{n=4}^{\infty} \frac{(n \!-\! 3)}{(n \!-\!2)!} \frac{(2 i x)^{n-2}}{a^n}
\Biggl\} \; .
\end{equation}
Employing this relation in (\ref{3fin}) and comparing with expression
(\ref{rfall}) for $\rho_{\rm falling}$ implies that we need the function
$F(x)$ to obey the relations,
\begin{eqnarray}
\int_0^{\infty} \!\!\!\! dx \, \frac{F(x)}{x^2} \, (1 \!+\! ix)^2 e^{-2ix}
+ {\rm c.c.} & \!\!\!= \!\!\!& \frac{3}{32 \pi^2} \; , \label{con1} \\
\int_0^{\infty} \!\!\!\! dx \, (ix)^{n-4} F(x)(1 \!+\! ix)^2 e^{-2ix}
+ {\rm c.c.} & \!\!\!=\!\!\! & \frac{(n \!+\!1) (n \!-\! 2)}{2^{n+1} n^2 \pi^2}
\, (n\!-\!4)! \quad \forall n \geq 4 \, . \qquad \label{con2}
\end{eqnarray}
It is useful to eliminate the factors of $i$ by defining real functions
$\alpha(x)$ and $\beta(x)$ as,
\begin{equation}
F(x) (1 \!+\! i x)^2 e^{-2ix} \equiv \alpha(x) + i \beta(x) \; .
\end{equation}
Then conditions (\ref{con1}-\ref{con2}) can be rewritten as,
\begin{eqnarray}
\int_0^{\infty} \!\!\! dx \, x^{-2} \alpha(x) & = & \frac{3}{64
\pi^2} \; , \label{new1} \\
\int_0^{\infty} \!\!\! dx \, x^{2m} \alpha(x) & = & \frac{(-1)^m (2m
\!+\!5) (m\!+\!1)}{2^{2m+7} \pi^2 (m \!+\! 2)^2} \times (2m)! \qquad
\forall m \geq 0 \; , \label{new2} \qquad \\
\int_0^{\infty} \!\!\! dx \, x^{2m+1} \beta(x) & = &
\frac{(-1)^{m+1} (m\!+\!3) (2m \!+\!3)}{2^{2m+6} \pi^2 (2m
\!+\!5)^2} \times (2m \!+\!1)! \qquad \forall m \geq 0 \; .
\label{new3} \qquad
\end{eqnarray}

Let us begin with (\ref{new3}). We can eliminate the factors of $2$
and $\pi$ by defining,
\begin{equation}
\beta(x) \equiv \frac{b(2x)}{32 \pi^2} \; ,
\end{equation}
and making the change of variable $y = 2x$. This implies,
\begin{eqnarray}
\int_0^{\infty} \!\! dy \, y^{2m + 1} b(y) &=& (-1)^{m+1} (2m \!+\!
1)! \, \frac{(2m \!+\! 6) (2m \!+\! 3)}{(2m \!+\! 5)^2} \; , \\
& = & (-1)^{m+1} (2m \!+\! 1)! \Biggl\{1 - \frac1{2m \!+\! 5} -
\frac2{(2m \!+\! 5)^2} \Biggr\} \; . \qquad
\end{eqnarray}
Now suppose we have found a function $b_1(y)$ which obeys,
\begin{equation}
\int_0^{\infty} \!\! dy \, y^{2m+1} b_1(y) = (-1)^{m+1} (2m \!+\!
1)! \; . \label{b1}
\end{equation}
We can employ it to construct functions $b_2(y)$ and $b_3(y)$ which
will add one and two factors of $1/(2m + 5)$, respectively,
\begin{eqnarray}
b_2(y) & \equiv & y^3 \int_y^{\infty} \!\! dz \, \frac{b_1(z)}{z^4}
\; , \label{b2} \\
b_3(y) & \equiv & y^3 \int_y^{\infty} \!\! dz \,
\frac{b_2(z)}{z^4} = y^3 \int_y^{\infty} \!\! dz \,
\frac{b_1(z)}{z^4} \, \ln\Bigl(\frac{z}{y}\Bigr) \; . \qquad
\label{b3}
\end{eqnarray}
Changing the order of integration shows that $b_2(y)$ has the
desired property,
\begin{eqnarray}
\int_0^{\infty} \!\! dy \, y^{2m+1} b_2(y) & = & \int_0^{\infty}
\!\! dy \, y^{2m+4} \int_y^{\infty} \!\! dz \, \frac{b_1(z)}{z^4} \;
, \\
& = & \int_0^{\infty} \!\! dz \, \frac{b_1(z)}{z^4} \int_0^{z} \!\!
dy \, y^{2m + 4} \; , \\
& = & \frac{(-1)^{m+1} (2m \!+\! 1)!}{2m \!+\! 5} \; .
\end{eqnarray}
Of course the same manipulations show that $b_3(y)$ has two powers
of $1/(2m+5)$. So if we can find $b_1(y)$ to enforce (\ref{b1}) then
we can construct $b_2(y)$ according to (\ref{b2}) and $b_3(y)$
according to (\ref{b3}) to give the function $\beta(x)$,
\begin{equation}
\beta(x) = \frac1{32 \pi^2} \Bigl[b_1(2x) \!-\!  b_2(2x) \!-\! 2
b_3(2x)\Bigr] \; .
\end{equation}

A solution for $b_1(y)$ seems to be just $\cos(y)$, provided we use
a convergence factor to make sense of the integral,
\begin{eqnarray}
\int_0^{\infty} \!\! dy \, e^{-\epsilon y} y^{2m+1} \cos(y) & = &
\Bigl(-\frac{\partial}{\partial \epsilon}\Bigr)^{2m+1}
\int_0^{\infty} \!\! dy \, e^{-\epsilon y} \cos(y) \; , \\
& = & \Bigl(-\frac{\partial}{\partial \epsilon}\Bigr)^{2m+1} \frac12
\Biggl\{ \frac1{\epsilon \!-\! i} + \frac1{\epsilon \!+\! i}\Biggr\}
\; , \\
& = & (2m \!+\! 1)! \frac12 \Biggl\{ \Bigl(\frac1{\epsilon \!-\!
i}\Bigr)^{2m+2} + \Bigl(\frac1{\epsilon \!+\! i}\Bigr)^{2m+2}
\Biggr\} \; . \qquad
\end{eqnarray}
Taking the limit $\epsilon \rightarrow 0^+$ gives the desired
relation,
\begin{equation}
\lim_{\epsilon \rightarrow 0^+} \int_0^{\infty} \!\! dy \,
e^{-\epsilon y} y^{2m+1} \cos(y) = (-1)^{m+1} (2m \!+\! 1)! \; .
\end{equation}
With a few partial integrations we can even express the function
$b_2(y)$ as a sine integral,
\begin{equation}
b_2(y) = y^3 \Biggl[ \frac{\cos(y)}{3 y^3} - \frac{\sin(y)}{6 y^2} -
\frac{\cos(y)}{6 y} - \frac16 {\rm Si}(y)\Biggr] \; .
\end{equation}
No similar expression can be obtained for $b_3(y)$.

The same pattern is followed in finding a function $\alpha(x)$ which
obeys (\ref{new2}). We first extract the factors of $2$ and $\pi$,
\begin{equation}
\alpha(x) = \frac{a(2x)}{32 \pi^2} \; ,
\end{equation}
which implies,
\begin{eqnarray}
\int_0^{\infty} \!\! dy \, y^{2m} a(y) & = & (-1)^m (2m)! \,
\frac{(2m \!+\! 5)(m \!+\!1)}{2 (m \!+\! 2)^2} \; , \\
& = & (-1)^m (2m)! \Biggl\{1 - \frac1{2 (m \!+\! 2)} - \frac1{2 (m
\!+\! 2)^2} \Biggr\} \; . \qquad
\end{eqnarray}
Hence we seek a function $a_1(y)$ with the property,
\begin{equation}
\int_0^{\infty} \!\! dy \, y^{2m} a_1(y) = (-1)^m (2m)! \; .
\end{equation}
From $a_1(y)$ we can construct $a_2(y)$ and $a_3(y)$ to insert
factors of $1/(2m +4)$ and $1/(2m+4)^2$, respectively,
\begin{eqnarray}
a_2(y) & = & y^3 \int_y^{\infty} \!\! dz \, \frac{a_1(z)}{z^4} \; ,
\\
a_3(y) & = & y^3 \int_y^{\infty} \!\! dz \, \frac{a_2(z)}{z^4} = y^3
\int_y^{\infty} \!\! dz \, \frac{a_1(z)}{z^4} \,
\ln\Bigl(\frac{z}{y}\Bigr) \; . \qquad
\end{eqnarray}
The function $\alpha(x)$ is,
\begin{equation}
\alpha(x) = \frac1{32 \pi^2} \Bigl[a_1(2x) - a_2(2x) - 2
a_3(2x)\Bigr] \; .
\end{equation}

It is straightforward to see that the desired solution for $a_1(y)$
is $\sin(y)$,
\begin{eqnarray}
\lim_{\epsilon \rightarrow 0^+} \int_0^{\infty} \!\! dy \,
e^{-\epsilon y} y^{2m} \sin(y) & = & \lim_{\epsilon \rightarrow 0^+}
\Bigl(\frac{\partial}{\partial \epsilon}\Bigr)^{2m} \frac1{2i}
\Biggl\{ \frac1{\epsilon \!-\! i} - \frac1{\epsilon \!+\! i}\Biggr\}
\; , \\
& = & (-1)^m (2m)! \; .
\end{eqnarray}
The function $a_2(y)$ can be expressed as a cosine integral,
\begin{equation}
a_2(y) = y^3 \Biggl\{ \frac{\sin(y)}{3 y^3} + \frac{\cos(y)}{6 y^2}
- \frac{\sin(y)}{6y} + \frac16 {\rm Ci}(y)\Biggr\} \; .
\end{equation}
It remains to note that relation (\ref{new1}) follows from analytic
continuation of (\ref{new2}) that we have just solved. First write
(\ref{new2}) in a form that makes sense for arbitrary $m$,
\begin{equation}
\frac{(-1)^m (2m \!+\! 5) (m \!+\! 1)}{2^{2m + 7} \pi^2 (m \!+\!
2)^2} \times (2m)! = \frac{e^{i m \pi} (2m \!+\! 5) (m \!+\! 1)}{
2^{2m +7} \pi^2 (m \!+\! 2)^2} \times \Gamma(2m \!+\! 1) \; .
\end{equation}
Then set $m = -1 + \epsilon$ and take the limit as $\epsilon$
approaches zero,
\begin{equation}
\lim_{\epsilon \rightarrow 0} \frac{e^{i (-1 + \epsilon) \pi} (3
\!+\! 2\epsilon) \epsilon}{2^{5 + 2\epsilon} \pi^2 (1 \!+\!
\epsilon)^2 } \times \Gamma(-1 \!+\! 2 \epsilon) = \frac{3}{64
\pi^2} \; .
\end{equation}

Assembling the various results of this section gives the following
final expression for the kernel function $F(k/H)$ of the state
corrections (\ref{+cor}-\ref{-cor}),
\begin{equation}
F(x) = \frac{i e^{2i x}}{32 \pi^2 (1 \!+\! ix)^2}
\Biggl\{e^{-2ix} 
- x^3 \!\! \int_{x}^{\infty} \frac{dz}{z^4}\,
e^{-2iz} - 2 x^3 \!\! \int_x^{\infty} \frac{dz}{z^4}\,
\ln\Bigl(\frac{z}{x}\Bigr) e^{-2 i z} \Biggr\} \; . \label{Fans}
\end{equation}

\section{Conclusions}

We have verified the conjecture \cite{OW} that the exponentially
redshifting parts (\ref{rfall}-\ref{pfall}) of the two loop energy
density and pressure of $\lambda \varphi^4$ theory on de Sitter
background can be completely absorbed into a redefinition of the
initial state. Our technique was to explicitly construct the
corrections (\ref{+cor}-\ref{-cor}), with the kernel function
$F(k/H)$ given in expression (\ref{Fans}). It might be worried
that we have only established the possibility of making this
modification of the initial state, not the necessity. However,
note that the parts of the free vacuum stress tensor we have
absorbed are not only exponentially falling, they also diverge on
the initial value surface. There is no alternative to absorbing
these terms initially, and making all time derivatives of the
stress tensor regular at least requires that the asymptotically
large powers of $1/a$ should be canceled.

It seems at least possible to give our state correction an elegant
interpretation. That would be to regard it as the finite remainder
of the $\lambda \phi^2$ correction that must come from the conformal
counterterm (\ref{dxi}). The idea is that a nonzero conformal
coupling $\delta \xi$ will change the mode functions $u(\eta,k)$
from (\ref{udef}) to,
\begin{equation}
u(\eta,k) \longrightarrow \sqrt{\frac{\pi}{4H}} \, a^{-\frac{D-1}2}
H^{(1)}_{\nu}\Bigl(\frac{k}{H a}\Bigr) \qquad {\rm with} \qquad
\nu^2 = \Bigl(\frac{D\!-\!1}2\Bigr)^2 - D (D\!-\!1) \delta \xi \; .
\end{equation}
Because the conformal counterterm changes only the quadratic part of
the Lagrangian density, the wave functional must still have the form
(\ref{freestate}) but with the new mode functions. Because $\delta
\xi$ is of order $\lambda$ one would expand the mode functions,
keeping only the first order correction for our current purposes.

The obvious problem with the interpretation we have just offered is,
{\it what becomes of the divergent part of} $\delta \xi$? We think a
possible answer is that there is also a correction of the form
$\lambda \phi^4$ which can contribute if two of the fields are taken
up by a coincident propagator and the other two connect to the
stress tensor at $x^{\mu}$. It then seems possible that the
divergence in the coincident propagator cancels against the
divergent part of $\delta \xi$, leaving the finite state correction
we have found. More work needs to be done to check this possibility.

We are obviously just at the beginning of systematically studying
and exploiting initial state corrections. One obvious application is
to cancel the surface terms that have been when two loop diagrams
are simplified by a partial integration \cite{SQED2}. Far from
simply being a complication, these surface terms would actually
lead, at higher orders, to new ultraviolet divergences which could
not be canceled by the usual volume counterterms.\footnote{We thank
A. Roura for pointing this out in the context of $\lambda \varphi^4$
theory.} Another important application will be to make the evolution
equations for quantum corrections to the mode function reliable at
finite times so that momentum dependent but temporally constant
changes in the normalization of mode functions can be reliably
determined \cite{KW2}. The possibility for observable tilts in the
power spectrum of primordial perturbations has already been noted
\cite{KW2}.

\vspace{1cm}

\centerline{\bf Acknowledgements} We have benefited from
discussions on this topic with T. Prokopec, A. Roura and N. C.
Tsamis. This work was partially supported by T\"UB\.ITAK projects
108T009, 107T896 and B\.IDEB 2221, by NSF grant PHY-0653085, and
by the Institute for Fundamental Theory at the University of
Florida.


\begin{thebibliography}{99}

\bibitem{IZ} C. Itzykson and J. B. Zuber, {\it Quantum Field Theory}
(McGraw-Hill, New York, 1980).

\bibitem{Weinberg} S. Weinberg, {\it The Quantum Theory of Fields},
Vol. II (Cambridge University Press, 1996).

\bibitem{JS} J. Schwinger, J. Math. Phys. {\bf 2} (1961) 407.

\bibitem{M} K. T. Mahanthappa, Phys. Rev. {\bf 126} (1962) 329;
P. M. Bakshi and K. T. Mahanthappa, J. Math. Phys. {\bf 4} (1963) 1;
J. Math. Phys. {\bf 4} (1963) 12.

\bibitem{K} L. V. Keldysh, Sov. Phys. JETP {\bf 20} (1965) 1018.

\bibitem{FW} L. H. Ford and R. P. Woodard, Class. Quant. Grav. {\bf 22}
(2005) 1637, gr-qc/0411003.

\bibitem{Hartle} J. B. Hartle and S. W. Hawking, Phys. Rev. {\bf D28} (1983)
2960.

\bibitem{Maldacena} J. Maldacena, JHEP {\bf 0305} (2003) 013, astro-ph/0210603.

\bibitem{Polyakov} A. M Polyakov, Nucl. Phys. {\bf B797} (2008) 199,
arXiv:0709.2899.

\bibitem{BD} N. D. Birrell and P. C. W. Davies, {\it Quantum Field Theory
on Curved Space} (Cambridge University Press, 1982).

\bibitem{AF} B. Allen and A. Folacci, Phys. Rev. {\bf D35} (1987) 3771.

\bibitem{TW1} N. C. Tsamis and R. P. Woodard, Annals Phys. {\bf 238} (1995) 1;
Phys. Lett. {\bf B426} (1998) 21, hep-ph/9710466.

\bibitem{OW} V. K. Onemli and R. P. Woodard, Class. Quant. Grav. {\bf 19}
(2002) 4607, gr-qc/0204065; Phys. Rev. {\bf D70} (2004) 107301, gr-qc/0406098.

\bibitem{SQED1} T. Prokopec, O. Tornkvist and R. P. Woodard, Phys. Rev. Lett.
{\bf 89} (2002) 101301, astro-ph/0205331; Annals Phys. {\bf 303} (2003) 251,
gr-qc/0205130; T. Prokopec and R. P. Woodard, Annals Phys. {\bf 312} (2004)
1, gr-qc/0310056.

\bibitem{Yukawa} T. Prokopec and R. P. Woodard, JHEP {\bf 0310} (2003) 059,
astro-ph/0309593; S. P. Miao and R. P. Woodard, Phys. Rev. {\bf D74} (2006)
044019, gr-qc/0602110.

\bibitem{QGDirac} S. P. Miao and R. P. Woodard, Class. Quant. Grav. {\bf 23}
(2006) 1721, gr-qc/0511140; Phys. Rev. {\bf D74} (2006) 024021, gr-qc/0603135;
Class. Quant. Grav. {\bf 25} (2008) 145009, arXiv:0803.2377.

\bibitem{SQED2} T.~Prokopec, N.~C.~Tsamis and R.~P.~Woodard, Class. Quant.
Grav. {\bf 24} (2007) 201, gr-qc/0607094; Annals Phys. {\bf 323} (2008)
1324, arXiv:0707.0847; Phys. Rev. {\bf D78} (2008) 043523, arXiv:0802.3673.

\bibitem{KO} E. O. Kahya and V. K. Onemli, Phys. Rev. {\bf D76} (2007)
043512, gr-qc/0612026.

\bibitem{TW2} N. C. Tsamis and R. P. Woodard, Class. Quant. Grav. {\bf 26}
(2009) 105006, arXiv:0807.5006.

\bibitem{KW1} E. O. Kahya and R. P. Woodard, Phys. Rev. {\bf D72} (2005)
104001, gr-qc/0508015; Phys. Rev. {\bf D74} (2006) 084012, gr-qc/0608049.

\bibitem{DW} L. D. Duffy and R. P. Woodard, Phys. Rev. {\bf D72} (2005)
024023, hep-ph/0505156.

\bibitem{KW2} E. O. Kahya and R. P. Woodard, Phys. Rev. {\bf D76} (2007)
124005, arXiv:0709.0536; Phys. Rev. {\bf D77} (2008) 084012, arXiv:0710.5282.

\bibitem{BOW} T. Brunier, V. K. Onemli and R. P. Woodard, Class. Quant. Grav.
{\bf 22} (2005) 59, gr-qc/0408080.

\bibitem{RGF} R. P. Woodard, Phys. Rev. Lett. {\bf 101} (2008) 081301,
arXiv:0805.3089.
\bibitem{refs} N. C. Tsamis and R. P. Woodard, arXiv:0904.2368; Phys.
Rev. D {\bf 78}, 028501 (2008), arXiv:0708.2004; Y. Wang,
arXiv:0904.2218; Phys. Rev. D {\bf 77}, 123525 (2008),
arXiv:0803.4295; arXiv:0712.0041; JCAP {\bf 0805} (2008) 021,
arXiv:0710.3885; G. Leon and E. N. Saridakis, arXiv:0904.1577; T.
M. Janssen, S. P. Miao, T. Prokopec and R. P. Woodard,
arXiv:0904.1151; J. Sadeghi, F. Milani and A. R. Amani,
arXiv:0904.0110; H. Zhang and H. Noh, arXiv:0904.0067; A.
Silvestri and M. Trodden, arXiv:0904.0024; I. Ya. Aref'eva, N. V.
Bulatov, L. V. Joukovskaya and S. Yu. Vernov, arXiv:0903.5264; E.
N. Saridakis, arXiv:0903.3840; arXiv:0902.3978; R. R. Caldwell and
M. Kamionkowski, arXiv:0903.0866; H. M. Sadjadi, arXiv:0902.2462;
JCAP {\bf 0702} (2007) 026, gr-qc/0701074; Y. Urakawa and T.
Tanaka, arXiv:0902.3209; arXiv:0904.4415; S. Sur, arXiv:0902.1186;
J. Wang and S.-P. Yang, arXiv:0901.1441; J. Wang, S.-W. Cui and
S.-P. Yang, arXiv:0901.1439; E. N. Saridakis, P. F. Gonzalez-Diaz
and C. L. Siguenza, arXiv:0901.1213; A. A. Sen, G. Gupta and S.
Das, arXiv:0901.0173; X.-M. Chen, Y.-G. Gong and E. N. Saridakis,
JCAP {\bf 0904} (2009) 001, arXiv:0812.1117; M. R. Setare, J.
Sadeghi and A. R. Amani, arXiv:0811.3343; Phys. Lett. B {\bf 660},
299 (2008), arXiv:0712.1873; M. R. Setare and E. N. Saridakis,
JCAP {\bf 0903} (2009) 002, arXiv:0811.4253; JCAP {\bf 0809}
(2008) 026, arXiv:0809.0114; arXiv:0807.3807; G. Cognola and S.
Zerbini, arXiv:0811.2714; W. Zhao, arXiv:0810.5506; Phys. Lett. B
{\bf 655}, 97 (2007), arXiv:0706.2211; Int. J. Mod. Phys. D {\bf
16}, 1735 (2007), gr-qc/0701136; M. Jamil, arXiv:0810.2896; M. P.
Lima, S. D. P. Vitenti and M. J. Reboucas, Phys. Lett. B {\bf
668}, 83 (2008), arXiv:0808.2467; Phys. Rev. D {\bf 77}, 083518
(2008), arXiv:0802.0706; F. Finelli, G. Marozzi, A. A.
Starobinsky, G. P. Vacca and G. Venturi, Phys. Rev. D {\bf 79},
044007 (2009), arXiv:0808.1786; S. Das, arXiv:0808.0826; J.
Santos, M. J. Reboucas and J. S. Alcaniz, arXiv:0807.2443;  T.
Janssen and T. Prokopec, arXiv:0807.0447; S. Sur and S. Das, JCAP
{\bf 0901} (2009) 007, arXiv:0806.4368; Y.-F. Cai and J. Wang,
Class. Quant. Grav. {\bf 25}, 165014 (2008), arXiv:0806.3890; H.
M. Sadjadi and N. Vadood, JCAP {\bf 0808} (2008) 036,
arXiv:0806.2767; S. H. Pereira and J. A. S. Lima, Phys. Lett. B
{\bf 669}, 266 (2008), arXiv:0806.0682; S. Weinberg, Phys. Rev. D
{\bf 78}, 063534 (2008), arXiv:0805.3781; M. R. Setare and J.
Sadeghi, Int. J. Theor. Phys. {\bf 47}, 3219, arXiv:0805.1117; S.
Unnikrishnan, Phys. Rev. D {\bf 78}, 063007 (2008),
arXiv:0805.0578; I. Ya. Aref'eva and A. S. Koshelev, JHEP {\bf
0809} (2008) 068, arXiv:0804.3570; Y. Wang, Phys. Rev. D {\bf 77},
123525 (2008), arXiv:0803.4295; J. F. Koksma and T. Prokopec,
Phys. Rev. D {\bf 78}, 023508 (2008), arXiv:0803.4000; S. Das and
N. Banerjee, Phys. Rev. D {\bf 78}, 043512 (2008),
arXiv:0803.3936; S. Wang, Y. Zhang and T.-Y. Xia, JCAP {\bf 0810}
(2008) 037, arXiv:0803.2760; T. Giannantonio, Y.-S. Song and K.
Koyama, Phys. Rev. D {\bf 78}, 044017 (2008), arXiv:0803.2238; P.
Mukherjee, M. Kunz, D. Parkinson and Y. Wang, Phys. Rev. D {\bf
78}, 083529 (2008), arXiv:0803.1616; G. Cognola and S. Zerbini,
Int. J. Theor. Phys. {\bf 47}, 3186 (2008), arXiv:0802.3967; M.
Novello and S. E. P. Bergliaffa, Phys. Rept. {\bf 463}, 127
(2008), arXiv:0802.1634; S. Unnikrishnan, H. K. Jassal and T. R.
Seshadri, Phys. Rev. D {\bf 78}, 123504 (2008), arXiv:0801.2017;
R. Aldrovandi, R. R. Cuzinatto and L. G. Medeiros, Eur. Phys. J. C
{\bf 58}, 483 (2008), arXiv:0801.0705; Y. Urakawa and K. Maeda,
Phys. Rev. D {\bf 78}, 064004 (2008), arXiv:0801.0126; M. R.
Setare, arXiv:0712.4004; N. Kaloper and S. Watson, Phys. Rev. D
{\bf 77}, 066002 (2008), arXiv:0712.1820; S. Tsujikawa, K. Uddin
and R. Tavakol, Phys. Rev. D {\bf 77}, 043007 (2008),
arXiv:0712.0082; H.-H. Xiong, T. Qiu, Y.-F. Cai and X. Zhang,
arXiv:0711.4469; A. A. Andrianov, F. Cannata, A. Y. Kamenshchik
and D. Regoli, JCAP {\bf 0802} (2008) 015, arXiv:0711.4300; Y.-F.
Cai, T. Qiu, R. Brandenberger, Y.-S. Piao and X. Zhang, JCAP {\bf
0803} (2008) 013, arXiv:0711.2187; M. R. Setare, Eur. Phys. J. C
{\bf 52}, 689 (2007), arXiv:0711.0524; P. Wu and H. W. Yu, Class.
Quant. Grav. {\bf 24}, 4661 (2007); JCAP {\bf 0710} (2007) 014,
arXiv:0710.1958; Phys. Lett. B {\bf 644}, 16 (2007),
gr-qc/0612055; A. K. Sanyal, arXiv:0710.3486; arXiv:0710.2450; V.
Faraoni, Phys. Rev. D {\bf 76}, 127501 (2007), arXiv:0710.1291; D.
Rapetti and S. W. Allen, MNRAS {\bf 388}, 1265 (2008),
arXiv:0710.0440; T. Qiu, Y.-F. Cai and X.-M. Zhang, Mod. Phys.
Lett. A {\bf 23}, 2787 (2008), arXiv:0710.0115; P. Martin-Moruno,
Phys. Lett. B {\bf 659}, 40 (2008), arXiv:0709.4410; S. Tsujikawa,
Phys. Rev. D {\bf 77}, 023507 (2008), arXiv:0709.1391; R. Lazkoz,
R. Maartens and E. Majerotto, J. Phys. Conf. Ser. {\bf 66}, 012057
(2007); W. Kim and E. J. Son, Mod. Phys. Lett. A {\bf 23}, 1079
(2008), arXiv:0708.1059; S.-F. Wu, A. Chatrabhuti, G.-H. Yang and
P.-M. Zhang, Phys. Lett. B {\bf 659}, 45 (2008), arXiv:0708.1038;
S. Yin, B. Wang, E. Abdalla and C.-Y. Lin, Phys. Rev. D {\bf 76},
124026 (2007), arXiv:0708.0992; T. Janssen and T. Prokopec,
arXiv:0707.3919; D. Seery, JCAP {\bf 0802} (2008) 006,
arXiv:0707.3378; JCAP {\bf 0711} (2007) 025, arXiv:0707.3377; L.
Joukovskaya, Phys. Rev. D {\bf 76}, 105007 (2007),
arXiv:0707.1545; S. K. Srivastava, arXiv:0707.1376; M. van der
Meulen and J. Smit, JCAP {\bf 0711} (2007) 023, arXiv:0707.0842;
J. C. C. de Souza and V. Faraoni, Class. Quant. Grav. {\bf 24},
3637 (2007); arXiv:0706.1223; H. Wei and S. N. Zhang, Phys. Rev. D
{\bf 76}, 063005 (2007), arXiv:0705.4002; Y. Gong and A. Wang,
Phys. Lett. B {\bf 652}, 63 (2007), arXiv:0705.0996; M. S.
Movahed, S. Baghram and S. Rahvar, Phys. Rev. D {\bf 76}, 044008
(2007), arXiv:0705.0889; S.-P. Miao, arXiv:0705.0767; L. Amendola
and S. Tsujikawa, Phys. Lett. B {\bf 660}, 125 (2008),
arXiv:0705.0396; L. Fernandez-Jambrina, Phys. Lett. B {\bf 656}, 9
(2007), arXiv:0704.3936; M. R. Setare, Phys. Lett. B {\bf 648},
329 (2007), arXiv:0704.3679; T. Naskar and J. Ward, Phys. Rev. D
{\bf 76}, 063514 (2007), arXiv:0704.3606; Y.-F. Cai, T. Qiu, Y.-S.
Piao, M. Li and X. Zhang, JHEP {\bf 0710} (2007) 071,
arXiv:0704.1090; J. Ren, X.-H. Meng and L. Zhao, Phys. Rev. D {\bf
76}, 043521 (2007), arXiv:0704.0672; A. Sheykhi, B. Wang and N.
Riazi, Phys. Rev. D {\bf 75}, 123513 (2007), arXiv:0704.0666; L.
Amendola, C. Charmousis and S. C. Davis, JCAP {\bf 0710} (2007)
004, arXiv:0704.0175; Y. Wang and P. Mukherjee, Phys. Rev. D {\bf
76}, 103533 (2007), astro-ph/0703780; Astrophys. J. {\bf 606}, 654
(2004), astro-ph/0312192; Astrophys. J. {\bf 650}, 1 (2006),
astro-ph/0604051; H. Zhang and Z.-H. Zhu, JCAP {\bf 0803} (2008)
007, astro-ph/0703245; J.-Q. Xia, Y.-F. Cai, T.-T. Qiu, G.-B. Zhao
and X. Zhang, Int. J. Mod. Phys. D {\bf 17}, 1229 (2008),
astro-ph/0703202; C. G. Boehmer and T. Harko, Gen. Rel. Grav. {\bf
39}, 757 (2007), gr-qc/0702078; I. Ya. Aref'eva, L. V. Joukovskaya
and S. Yu. Vernov, JHEP {\bf 0707} (2007) 087, hep-th/0701184; S.
Fay, R. Tavakol and S. Tsujikawa, Phys. Rev. D {\bf 75}, 063509
(2007), astro-ph/0701479; X. Zhang and F.-Q. Wu, Phys. Rev. D {\bf
76}, 023502 (2007), astro-ph/0701405; R. Lazkoz, G. Leon and I.
Quiros, Phys. Lett. B {\bf 649}, 103 (2007), astro-ph/0701353; A.
S. Koshelev, JHEP {\bf 0704} (2007) 029, hep-th/0701103; Y.-F.
Cai, M.-Z. Li, J.-X. Lu, Y.-S. Piao, T.-T. Qiu and X.-M. Zhang,
Phys. Lett. B {\bf 651}, 1 (2007), hep-th/0701016; L. Amendola, R.
Gannouji, D. Polarski and S. Tsujikawa, Phys. Rev. D {\bf 75},
083504 (2007), gr-qc/0612180; Y.-H. Wei, Mod. Phys. Lett. A {\bf
21}, 2845 (2006); H. Wei, N.-N. Tang and S. N. Zhang, Phys. Rev. D
{\bf 75}, 043009 (2007), astro-ph/0612746; S. Winitzki, Lect.
Notes Phys. {\bf 738}, 157 (2008), gr-qc/0612164; S. Nesseris and
L. Perivolaropoulos, JCAP {\bf 0702} (2007) 025, astro-ph/0612653;
F. Briscese, E. Elizalde, S. Nojiri and S. D. Odintsov, Phys.
Lett. B {\bf 646}, 105 (2007), hep-th/0612220; S. Yu. Vernov,
Theor. Math. Phys. {\bf 155}, 544 (2008), astro-ph/0612487; I. Ya.
Aref'eva and I. V. Volovich, Theor. Math. Phys. {\bf 155} 503
(2008), hep-th/0612098; H.-S. Zhang and Z.-H. Zhu, Phys. Rev. D
{\bf 75}, 023510 (2007), astro-ph/0611834; R. V. Buniy, S. D. H.
Hsu and B. M. Murray, Phys. Rev. D {\bf 74}, 063518 (2006),
hep-th/0606091; R. V. Buniy and S. D. H. Hsu, Phys. Lett. B {\bf
632}, 543 (2006), hep-th/0502203; A. Dolgov and D. N. Pelliccia,
Nucl. Phys. {\bf B734}, 208 (2006), hep-th/0502197; A. Vikman,
Phys. Rev. D {\bf 71}, 023515 (2005), astro-ph/0407107; S. D. H.
Hsu, A. Jenkins and M. B. Wise, Phys. Lett. B {\bf 597}, 270
(2004), astro-ph/0406043; A. Lue and G. D. Starkman, Phys. Rev. D
{\bf70}, 101501 (2004), astro-ph/0408246; S. M. Carroll, A. De
Felice and M. Trodden, Phys. Rev. D {\bf71}, 023525 (2005),
astro-ph/0408081; U. Alam, V. Sahni, T. D. Saini and A. A.
Starobinsky, MNRAS {\bf 354}, 275 (2004), astro-ph/0311364; R.
Lazkoz, S. Nesseris and L. Perivolaropoulos, JCAP {\bf 0511}
(2005) 010, astro-ph/0503230; S. Nesseris and L. Perivolaropoulos,
JCAP {\bf 0702} (2007) 025, astro-ph/0612653; JCAP {\bf 0701}
(2007) 018, astro-ph/0610092; Phys. Rev. D {\bf70}, 043531 (2004),
astro-ph/0401556; J. M. Cline, S. Y. Jeon and G. D. Moore, Phys.
Rev. D {\bf 70}, 043543 (2004), hep-ph/0311312; S. M. Carroll, M.
Hoffman and M. Trodden, Phys. Rev. D {\bf 68}, 023509 (2003),
astro-ph/0301273; R. R. Caldwell, Phys. Lett. B {\bf 545}, 23
(2002), astro-ph/9908168.


\bibitem{AAS} A. A. Starobinsky, ``Stochastic de Sitter (inflationary) stage
in the early universe,'' in {\it Field Theory, Quantum Gravity and Strings,}
ed. H. J. de Vega and N. Sanchez (Springer-Verlag, Berlin, 1986) pp. 107-126.

\bibitem{SY} A. A. Starobinsky and J. Yokoyama, Phys. Rev. {\bf D50} (1994)
6357, astro-ph/9407016.

\bibitem{TW3} N. C. Tsamis and R. P. Woodard, Nucl. Phys. {\bf B724} (2005)
295, gr-qc/0505115.

\bibitem{BA} B. Allen, Phys. Rev. {\bf D32} (1985) 3136.

\bibitem{TW4} N. C. Tsamis and R. P. Woodard, Class. Quant. Grav. {\bf 11}
(1994) 2969.

\bibitem{ITTW} J. Iliopoulos, T. N. Tomaras, N. C. Tsamis and R. P. Woodard,
Nucl. Phys. {\bf B534} (1998) 419, gr-qc/9801028.

\bibitem{JMPW} T. M. Janssen, S. P. Miao, T. Prokopec and R. P. Woodard,
Class. Quant. Grav. {\bf 25} (2008) 245013, arXiv:0808.2449.

\bibitem{revs} K. C. Chou, Z. B. Su, B. L. Hao and L. Yu, Phys. Rept. {\bf 118}
(1985) 1; R. D. Jordan, Phys. Rev. {\bf D33} (1986) 444; E. Calzetta and B.
L. Hu, Phys. Rev. {\bf D35} (1987) 495.

\end{thebibliography}
\end{document}